\documentclass{article}%
\usepackage{amssymb}
\usepackage{graphicx}
\usepackage{amsmath}
\usepackage{amsfonts}%
\providecommand{\U}[1]{\protect\rule{.1in}{.1in}}
\newtheorem{theorem}{Theorem}

\newtheorem{lemma}[theorem]{Lemma}

\begin{document}

\begin{center}
Quantum spin pump on a finite antiferromagnetic chain through bulk states

\bigskip

Nan-Hong Kuo$^{1}$, Sujit Sarkar$^{2}$, C. D. Hu$^{1,3\ast}$

$^{1}$Department of Physics, National Taiwan University, Taipei, Taiwan, R.O.C.

$^{2}$PoornaPrajna Institute of Scientific Research, 4 Sadashivanagar,
Bangalore-5600 80, India

$^{3}$Center for Theoretical Sciences, National Taiwan University, Taipei,
Taiwan, R.O.C.

\bigskip

Abstract
\end{center}

We studied the possibility of the spin pump in a S=1/2 antiferromagnetic
chain. The spin chain is mapped into a fermion system and bosonization is
utilized to transform the equation of motion to a sine-Gordon equation. The
sine-Gordon equation on a finite chain with different boundary conditions is
solved. Among numerous solutions, the static soliton is compatible with the
original physical system. By varying adiabatically a angle $\phi$\ in the
phase space composed of applied electric and magnetic fields, the spin states
change between the N\'{e}el state and dimer state and a quantized spin $S=1$
is transported by the bulk state from one end of the system to the other.

\bigskip

\noindent PACS: 75.10.Pq, 75.10.Jm, 03.65.Vf

\noindent Keywords: spin chain, sine-Gordon equation, quantum spin
transport\newpage

\paragraph{1. Introduction:}

An adiabatic quantum pump is a device that generates a dc current by a cyclic
variation of some system parameters, the variation being slow enough so that
the system remains close to the ground state throughout the pumping cycle.
After the pioneering work of Thouless[1] and Niu and Thoulesss[2], the quantum
adiabatic pumping physics gets more attention.\ It is applied to the systems
like open quantum dots[3-5], superconducting quantum wires[6,7], the Luttinger
quantum wire[8], the interacting quantum wire[9] and of course the spin sytems.

In recent years, spintronics become an exciting new field new field of
research. Various proposals of generating spin current have been studied.
Among them, an adiabatic spin pumping process is most interesting. Quantum
spin pumping physics probably has inspired by the phenomenal work of
Thouless[1], which is clearly related to the topological explanation of
quantum hall effect by Thouless $et.al$.[10]. However Halperin[11] pointed out
before that the quasi-one dimensional edge states played an important role in
quantum hall effect. Hatsugai[12,13] showed that the edge states indeed have
topological meaning and thus confirm their importance.\newline

Shindou[14] has shown that the origin of spin transport is due to the edge
state of the system. Fu, Kane and Mele[15] and Fu and Kane[16] studied the
similar problem. Among their contribution, they found that the edge state
crossing (Kramers degeneracy) is essential for spin pumping. So the
possibility of spin transport through the bulk states of the system is not
reveal from these studies and leave this bulk state spin transport as a open
problem.\newline

Here we mention very briefly the basic theme of spin transport in adiabatic
process. Suppose we consider a spin chain and constructed a parameter space
with $(h_{st},\Delta)=R(\cos{\phi},\sin{\phi})$ where $h_{st}$ is the applied
magnetic field $\Delta$ the dimer states bond strength. Fixing $R$ and varying
$\phi$ adiabatically in time, one can argue that a line integral of
$\mathbf{A}{_{n}}(\mathbf{K})=(i/2\pi)<n(\mathbf{K})|{{\nabla}_{\mathbf{K}}%
}|n(\mathbf{K})>$, where $n(\mathbf{K})$ is the Bloch function for the n-th
band, and $\mathbf{K}=(k,{\Delta},h_{st})$, on a closed loop yields exactly
$\pm1$ due to the singularity at the origin. In other words, $\mathbf{A}$ is
related to a fictitious magnetic field $\mathbf{B}{_{n}}(\mathbf{K})={{\nabla
}_{\mathbf{K}}}\times\mathbf{A}{_{n}}(\mathbf{K})$. One with the Stokes'
theorem, can express the line integral in terms of surface integral
($\int\mathbf{B}\cdot d\mathbf{S}$) where the integration is on two
dimensional closed surface enclosing the origin. This is exactly the
quantization of particle transport proposed by Thouless[1]. It is well known
to us that one can express spin chain problem into a spin less fermion problem
with Jordan-Wigner transformations. One can use this kind of adiabatic
variation of parameters as a tool of quantized spin transport. Shindou
considered the spin polarization $P_{s^{z}}=\frac{1}{N}\sum_{j=1}^{N}j{S_{j}%
}^{z}$ and divided it into two parts, bulk state part and edge state part and
he concluded that edge state part of spin polarization contribute to spin
transport. Fu and Kane[16] considered a similar system with an additional
interaction of spin-orbit coupling. They calculated the energy bands of the
bulk states and end (edge) states and were able to show clearly that whenever
there is Kramers degeneracy of end states, there is spin transport and it has
a $Z_{2}$ symmetry.\newline

We have already mentioned that in all previous studies of spin transport, the
contribution is coming from the edge states. Here, most probably first in the
literature, we raise the question, whether the edge states are indispensable
in spin transport?. We do the rigorous analytical exercises to complete the
search of this question. One can see during our analytical derivation that
spin transport is nothing but the transport of soliton in the system. The plan
of our paper is the following: In section (2), we present model Hamiltonian
and continuum field theoretical studies. Section (3) is for analysis of
sine-Gordon equation on a finite chain for different boundary conditions.
Section (4) is for the detail analysis for static soliton solution. Section
(5) is reserved for results and discussions.

\paragraph{2. Hamiltonian and Continuum field theoretical studies}

We consider a spin 1/2 chain of finite length, described by the Heisenberg
Hamiltonian similar to that of Shindou[14]. A controlled dimerization
amplitude and applied magnetic field are also present. The total Hamiltonian
has three parts:
\begin{equation}
H(t)=H_{0}+H_{\dim}+H_{st} \tag{1}%
\end{equation}
where%

\begin{equation}
H_{0}=J\sum\limits_{i=1}^{N}\mathbf{S}_{i}\cdot\mathbf{S}_{i+1}, \tag{2}%
\end{equation}%
\begin{equation}
H_{\dim}=\frac{\Delta(t)}{2}\sum\limits_{i=1}^{N}(-1)^{i}(S_{i}^{+}S_{i+1}%
^{-}+S_{i}^{-}S_{i+1}^{+}), \tag{3}%
\end{equation}
and
\begin{equation}
H_{st}=h_{st}(t)\sum\limits_{i}(-1)^{i}S_{i}^{z}. \tag{4}%
\end{equation}
$H_{\dim}$ is the bond alternation term which can be induced by applying an
electric field to the spin chain to alter the exchange interaction. It
introduces into the system the strength of dimerization $\Delta(t)$. $H_{st} $
is the coupling of the system to a staggered external field $h_{st}(t)$. The
time-dependent bond strength $\Delta(t)$ and staggered field $h_{st}(t)$ can
be varied adiabatically so as to create a parameter space for Berry phase. We
write $\Delta$ and $h_{st}$ as $(h_{st},\Delta)=R(\cos\phi,\sin\phi)$, with
$R$ fixed. Varying $\phi$ adiabatically, we expect spins to be transported. We
shall also argue that going through one cycle along the loop, there will be
quantized spin component transported from one end to the other.

The method of bosonization[17-20] has been used successfully to treat various
one-dimensional systems, including the spin chains. It is suitable for the
system we are considering. To this end, we first make the Jordon-Winger
transformation to represent spins by fermion field $f_{i}$ and $f_{i}^{+}$.
Then, the bosonizations of $f_{i}$ and $f_{i}^{+}$ will performed.
\begin{equation}
f_{j}=\exp(i\pi%
%TCIMACRO{\dsum \limits^{j-1}}%
%BeginExpansion
{\displaystyle\sum\limits^{j-1}}
%EndExpansion
S_{i}^{+}S_{i}^{-})S_{j}^{-}, \tag{5}%
\end{equation}%
\begin{equation}
f_{j}^{\dag}=S_{j}^{+}\exp(-i\pi%
%TCIMACRO{\dsum \limits^{j-1}}%
%BeginExpansion
{\displaystyle\sum\limits^{j-1}}
%EndExpansion
S_{i}^{+}S_{i}^{-}), \tag{6}%
\end{equation}
and
\begin{equation}
f_{j}\backsimeq R(x_{j})e^{ik_{F}x_{j}}+L(x_{j})e^{-ik_{F}x_{j}}, \tag{7}%
\end{equation}%
\begin{equation}
f_{j}^{\dag}\backsimeq R^{\dag}(x_{j})e^{-ik_{F}x_{j}}+L^{\dag}(x_{j}%
)e^{ik_{F}x_{j}}, \tag{8}%
\end{equation}
where
\begin{equation}
R(x)=\frac{1}{\sqrt{2\pi\alpha}}\eta_{1}e^{i[\theta_{+}(x)+\theta_{-}(x)]/2}
\tag{9}%
\end{equation}
and
\begin{equation}
L(x)=\frac{1}{\sqrt{2\pi\alpha}}\eta_{2}e^{i[-\theta_{+}(x)+\theta_{-}(x)]/2}
\tag{10}%
\end{equation}
are the slowly varying fields, and $\eta_{1}$ and $\eta_{2}$ are the Klein
factors. Here, $k_{F}$ is the Fermi wave vector and and $\alpha$ is the
lattice constant. For a half filled system we have $k_{F}=\pi/2\alpha$. In the
following derivations, we left out the details because they can be found in
many textbooks[17-20].
\begin{align}
S_{j}^{z}  &  =f_{j}^{+}f_{j}-\frac{1}{2}\tag{11}\\
&  =\frac{\partial_{x}\widehat{\theta}_{+}(x_{j})}{2\pi}-(-1)^{j}\frac{1}%
{\pi\alpha}\sin\widehat{\theta}_{+}(x_{j})\nonumber
\end{align}
and
\begin{align}
S_{i}^{+}S_{i+1}^{-}+S_{i}^{-}S_{i+1}^{+}  &  =f_{j}^{+}f_{j+1}+f_{j+1}%
^{+}f_{j}\nonumber\\
&  =-\alpha\lbrack4\pi\widehat{\Pi}^{2}+\frac{1}{4\pi}(\partial_{x}%
\widehat{\theta}_{+})^{2}]-(-1)^{j}\frac{1}{\pi\alpha}\cos\widehat{\theta}%
_{+}(x_{j}). \tag{12}%
\end{align}
Here $\theta_{+}=$ is bosonization phase and\ $\widehat{\Pi}(x)=-(1/4\pi
)\partial_{x}\theta_{-}(x)$ is the conjugate momentum of $\theta_{+}%
(x)$.\ Substituting eqs. (11) and (12) into eqs. (1-4), and dropping the
rapidly varying components such as $\sum\limits_{j}(-1)^{j}\cos\widehat
{\theta}_{+}(x_{j})$, we obtained%

\begin{equation}
H=\int dx\{v[\pi\eta\widehat{\Pi}^{2}+\frac{1}{4\pi\eta}(\partial_{x}%
\widehat{\theta}_{+})^{2}]-\frac{R}{\pi\alpha^{2}}\sin(\widehat{\theta}%
_{+}+\varphi)+\frac{J}{2\pi^{2}\alpha^{3}}\cos2\widehat{\theta}_{+}\} \tag{13}%
\end{equation}
where the velocity
\begin{equation}
v=J\sqrt{1+\frac{2}{\pi}} \tag{14}%
\end{equation}
and the quantum parameter
\begin{equation}
\eta=\frac{J}{v} \tag{15}%
\end{equation}
\ were discussed in ref. 17-20. Thus, we have the equation of motion
\begin{equation}
\partial_{t}^{2}\widehat{\theta}_{+}=v^{2}\partial_{x}^{2}\widehat{\theta}%
_{+}+\frac{2JR}{\alpha^{2}}\cos(\widehat{\theta}_{+}+\varphi)+\frac{2J^{2}%
}{\pi\alpha^{3}}\sin2\widehat{\theta}_{+} \tag{16}%
\end{equation}
The term of $\sin2\widehat{\theta}_{+}$ is irrelevant in the sense of
renormalization group analysis, so we consider only the part%

\begin{equation}
\partial_{\tau}^{2}\widehat{\theta}_{+}=\partial_{z}^{2}\widehat{\theta}%
_{+}+\cos(\widehat{\theta}_{+}+\varphi). \tag{17}%
\end{equation}
where we have change variables: $z=\sqrt{2JR}x/\nu\alpha$ and $\tau=\sqrt
{2JR}t/\alpha$. It is similar to the standard sine-Gordon equation
\begin{equation}
\partial_{\tau}^{2}\widehat{\theta}_{+}-\partial_{z}^{2}\widehat{\theta}%
_{+}+\sin\widehat{\theta}_{+}=0 \tag{18}%
\end{equation}
which has been well-studied. However, for our purpose which is to study the
spin transport, we will solve it on a chain of finite length on which the
phase $\phi$\ is no longer a trivial\ constant but introduces new meaning to
the solution. This way, one can recognize the motion of spins from one end to
the other.

\paragraph{3. Analysis of sine-Gordon equation on a finite chain}

We shall analyze eq. (18) first. The result can be applied to eq. (17).
Equation (18) has many kinds of solutions. The traveling-wave solutions, such
as $\arctan[\exp(\gamma(z-v\tau))]$ is not suitable for our purpose because
they cannot meet fixed boundary conditions. For the finite-length systems, we
consider the so called \textquotedblright separable
solutions\textquotedblright\lbrack21-23].%

\begin{equation}
\phi(z,\tau)=4\arctan(A\frac{f(\beta z)}{g(\Omega\tau)}). \tag{19}%
\end{equation}
and $f(\beta z)$ and $g(\Omega\tau)$ must satisfy the following equations:.%

\begin{equation}
(\partial_{z}f)^{2}=(\frac{1}{\beta^{2}})[-\kappa A^{2}f^{4}+\mu f^{2}%
+(\frac{\lambda}{A^{2}})] \tag{20}%
\end{equation}
and
\begin{equation}
(\partial_{\tau}g)^{2}=(\frac{1}{\Omega^{2}})[-\lambda g^{4}+(\mu
-1)g^{2}+\kappa]. \tag{21}%
\end{equation}
with the requirements $\mu^{2}+4\kappa\lambda\geq0$ and $(\mu-1)^{2}%
+4\kappa\lambda\geq0$.$\ A,\ \mu,\ \kappa,\ \lambda\ \beta\ $and$\ \Omega
\ $are\ mutually related constants. We will show how they are determined in a
while. First, we would like to put forward the observation that $f(\beta z)$
and $g(\Omega\tau)$ satisfying eqs. (20) and (21) are Jacobi elliptic
functions (JEF)[24]. Jacobi elliptic functions are defined as the following:
\begin{equation}
u=%
%TCIMACRO{\dint \nolimits_{0}^{sn(u)}}%
%BeginExpansion
{\displaystyle\int\nolimits_{0}^{sn(u)}}
%EndExpansion
\frac{dt}{\sqrt{(1-t^{2})(1-k^{2}t^{2})}}. \tag{22}%
\end{equation}
where $sn(u)$ is one of the JEFs and $k$\ is a constant in the range [0,1].
The second JEF is $cn(u)$ where $sn^{2}(u)+cn^{2}(u)=1$.\ There are more JEFs.
They can be found in Appendix A. The ones we are going to encounter are
$sc(u)=sn(u)/cn(u)$\ and $dn^{2}(u)=1-k^{2}sn^{2}(u)$.\ Both $f$\ and $g$\ are
JEFs and their constants are denoted by $k_{f}$\ and $k_{g}$.\ $\mu$%
,\ $\kappa$\ and $\lambda$\ are constants determined by $k_{f}$\ and\ $k_{g}$.
The relations are different for different Jacobi elliptic functions.

Here we give an example of $f(\beta z)=cn(\beta(z-z_{0}))$\ and\ $g(\Omega
\tau)=cn(\Omega\tau)$ where $z_{0}$\ is a constant.\ With the equation for
$cn(u)$ (see Table I in Appendix A)%
\begin{equation}
(\partial_{u}cn(u))^{2}=(1-u^{2})(1-k^{2}+k^{2}u^{2}), \tag{23}%
\end{equation}
we found from comparison with eq. (20) that $\kappa A^{2}=k_{f}^{2}$,
$\mu=2k_{f}^{2}-1$ and $\lambda/A^{2}=1-k_{f}^{2}$. As a result, we get
\begin{equation}
k_{f}=\frac{A^{2}}{1+A^{2}}+\frac{A^{2}}{\beta^{2}(1+A^{2})^{2}}. \tag{24}%
\end{equation}
and,\
\begin{equation}
k_{g}=\frac{A^{2}}{1+A^{2}}-\frac{A^{2}}{\Omega^{2}(1+A^{2})^{2}}. \tag{25}%
\end{equation}
where\
\begin{equation}
\Omega^{2}=\beta^{2}+\frac{1-A^{2}}{1+A^{2}}. \tag{26}%
\end{equation}
If we choose the fixed boundary condition $\theta_{+}(z=0)=\theta_{+}(z=L)=0$
with\ $L$\ being the length of the system, then we will have
\begin{equation}
\beta L=4lK(k_{f}) \tag{27}%
\end{equation}
with
\begin{equation}
K=%
%TCIMACRO{\dint \nolimits_{0}^{1}}%
%BeginExpansion
{\displaystyle\int\nolimits_{0}^{1}}
%EndExpansion
\frac{dt}{\sqrt{(1-t^{2})(1-k^{2}t^{2})}} \tag{28}%
\end{equation}
being the complete elliptic integral of the first kind, $z_{0}=L/4l$ and
$l$\ is an integer.\ 

Not all the combinations of JEFs can satisfy the sine-Gordon equation. A table
of the differential equations for all the JEFs is given in Appendix A. We will
discuss the solutions of the sine-Gordon equation in eq. (17) on a finite
system under various boundary conditions. Although finite-length solutions are
well-known, different boundary conditions and the presence of $\phi$ will
impose restrictions on the solutions and infuse physical meaning to the wave forms.

\subparagraph{Case 1: Periodic boundary condition $\widehat{\theta}_{+}%
(z,\tau)=\widehat{\theta}_{+}(z+L,\tau)$}

The first boundary condition coming to mind is the periodic boundary
condition. There are many combinations of JEFs that can satisfy the periodic
boundary condition. Here are two examples.

1a:
\begin{equation}
\widehat{\theta}_{+}(z,\tau)=\frac{\pi}{2}-\varphi+4\arctan\{Acn[\beta
(z-z_{0});k_{f}]cn[\Omega\tau;k_{g}]\} \tag{29}%
\end{equation}
where $\beta L=4lK(k_{f})$ and

1b:
\begin{equation}
\widehat{\theta}_{+}(z,\tau)=\frac{\pi}{2}-\varphi+4\arctan\{Asc[\beta
(z-z_{0});k_{f}]dn[\Omega\tau;k_{g}]\} \tag{30}%
\end{equation}
where $\beta L=2lK(k_{f})$ and $z_{0}$\ is a arbitrary constant. For this
boundary condition, there is no spin transport if one varies the parameter
$\phi$ adiabatically. The reason is quite simple. Increasing $\phi$ only give
a constant change to $\theta_{+}(z)$\ everywhere. Thus the fermion field
operators on every site from Jordan-Wigner transformation acquire a constant
phase and the spins remain the same.

\subparagraph{Case 2: Fixed boundary condition $\widehat{\theta}_{+}%
(z=0,\tau)=\widehat{\theta}_{+}(z=L,\tau)=0$}

It seems that we can have the solutions like
\begin{equation}
\widehat{\theta}_{+}(z,\tau)=\frac{\pi}{2}-\varphi+4\arctan\{Acn[\beta
(z-z_{0});k_{f}]cn[\Omega\tau;k_{g}]\} \tag{31}%
\end{equation}
where $\beta L=2lK(k_{f})$. However, the presence of the adiabatic change term
$\pi/2-\varphi$ in front requires that $cn[\beta(z-z_{0});k_{f}]$ to be
finite. Then the function $cn[\Omega\tau;k_{g}]$ makes the inverse tangent
function varying with time and hence, the forms in solution (32) can not
satisfy the fixed boundary condition.

\subparagraph{Case 3: Free end boundary condition $(\partial\widehat{\theta
}_{+}(z,\tau)/\partial z)|_{z=0}=(\partial\widehat{\theta}_{+}(z,\tau
)/\partial z)|_{z=L}=0$}

The solution is
\begin{equation}
\widehat{\theta}_{+}(z,\tau)=\frac{\pi}{2}-\varphi+4\arctan\{Adn[\beta
(z-z_{0});k_{f}]sn[\Omega\tau;k_{g}]\} \tag{32}%
\end{equation}
where $\beta L=2K_{f}$ and $\beta z_{0}=K_{f}$. The energy is equal to
$16\beta E(K)$ where
\begin{equation}
E(K)=%
%TCIMACRO{\dint \limits_{0}^{\pi/2}}%
%BeginExpansion
{\displaystyle\int\limits_{0}^{\pi/2}}
%EndExpansion
\frac{\sqrt{1-k^{2}t^{2}}}{\sqrt{1-t^{2}}}dt \tag{33}%
\end{equation}
is the complete elliptic integral of the second kind. This solution cannot
provide the system with spin transport for the same reason as that in case 1.

\paragraph{4. Detailed analysis of the static soliton case}

It is most interesting to study the solution of the static soliton of eq.
(17). Let us first consider the boundary condition
\begin{align}
\theta_{+}(z  &  =0)=0,\tag{34}\\
\theta_{+}(z  &  =L)=2\pi.\nonumber
\end{align}
The phase difference $2\pi$\ implies that the fermion field or the spins have
same boundary conditions at both ends and hence, a common case for a finite
spin chain. On the other hand, it is a fixed boundary condition of
$\widehat{\theta}_{+}$. Therefore, different values of $\phi$\ will induce
distinct solutions. The soliton has the form
\begin{equation}
\widehat{\theta}_{+}(z,\tau)=\frac{\pi}{2}-\varphi+4\arctan\{Asc[\beta
(z-z_{0});k_{f}]dn[\Omega\tau;k_{g}]\} \tag{35}%
\end{equation}
where
\begin{equation}
k_{f}^{2}=1-A^{2}+\frac{A^{2}}{\beta^{2}(1-A^{2})}, \tag{36}%
\end{equation}%
\begin{equation}
k_{g}^{2}=1-\frac{1}{A^{2}}+\frac{1}{\Omega^{2}(1-A^{2})}, \tag{37}%
\end{equation}%
\begin{equation}
\Omega=A\beta, \tag{38}%
\end{equation}
and
\begin{equation}
\beta L=K(k_{f}). \tag{39}%
\end{equation}
Equations (36-38) can be derived by substituting eq. (35) into eq. (17) and
eq. (39) comes from the boundary conditions in eq. (34). We seek the static
solution because it can always satisfy above boundary conditions. In this
case, we require $k_{g}=0$, $dn(\Omega\tau,k_{g}=0)=1$ and $A$\ takes a
special value $A_{th}$.\ In view of eqs. (36) and (37), the static soliton is
\begin{equation}
\widehat{\theta}_{+}(z,\tau)=\frac{\pi}{2}-\varphi+4\arctan\{A_{th}%
sc[\beta(z-z_{0});k_{f}]\} \tag{40}%
\end{equation}
with
\begin{equation}
\beta=1/(1-A_{th}^{2}) \tag{41}%
\end{equation}
and
\begin{equation}
k_{f}=\sqrt{1-A_{th}^{4}} \tag{42}%
\end{equation}
\ The boundary condition at $z=0$ requires that
\begin{equation}
\tan(\frac{\phi}{4}-\frac{\pi}{8})=A_{th}sc(-\beta z_{0}) \tag{43}%
\end{equation}
which determines $z_{0}$. For the boundary condition at $z=L$, we have derived
the following lemma:

For the solution in eq. (40) with
\begin{equation}
\beta L=K(k_{f}=\sqrt{1-A_{th}^{4}})=\int_{0}^{1}\frac{dt}{\sqrt
{(1-t^{2})[1-(1-A_{th}^{4})t^{2}]}}, \tag{44}%
\end{equation}
the difference of $\widehat{\theta}_{+}(z=0)$\ and\ $\widehat{\theta}%
_{+}(z=L)$\ is always equal to $2\pi$.

The derivation is given in Appendix B.\ Eqs. (40) and (44) are the main result
of this paper. Equation (44) can be generalized as
\begin{equation}
\beta L=lK \tag{45}%
\end{equation}
where $l$ is any nonzero integer. As The larger the magnitude of $l$, the
higher the energy.

In Fig. 1 $A_{th}$\ evaluated with eq. (44) is plotted. It shows that $A_{th}
$\ decreases rapidly with increasing $L$. The magnitude of $A_{th}$\ is
closely related to the wave form of $\widehat{\theta}_{+}$. A small $A_{th}%
$\ results in a steep change in $\widehat{\theta}_{+}$., or a sharp domain
wall. As it will be shown later, it related to the quantum spin transport.

The energy of the static soliton can be calculated with%

\begin{equation}
\mathcal{E}=\int_{0}^{L}dz[\frac{1}{2}(\frac{\partial\widehat{\theta}_{+}%
}{\partial\tau})^{2}+\frac{1}{2}(\frac{\partial\widehat{\theta}_{+}}{\partial
z})^{2}-\sin(\widehat{\theta}_{+}+\phi)]. \tag{46}%
\end{equation}
It can be shown that
\begin{align}
&  \frac{1}{2}(\frac{\partial\widehat{\theta}_{+}}{\partial z})^{2}%
+\sin(\widehat{\theta}_{+}+\phi)-\frac{1}{2}(\frac{\partial\theta_{+}%
}{\partial\tau})^{2}|_{z=z_{0}}-\sin(\widehat{\theta}_{+}+\phi)|_{z=z_{0}%
}\tag{47}\\
&  =\frac{1}{2}(\frac{\partial\widehat{\theta}_{+}}{\partial z})^{2}%
+\sin(\widehat{\theta}_{+}+\phi)-\frac{8A_{th}^{2}}{(1-A_{th}^{2})^{2}%
}-1=0\nonumber
\end{align}
Hence, eq. (45) becomes
\begin{equation}
\mathcal{E}=\int_{0}^{L}dz[1+\frac{8A_{th}^{2}}{(1-A_{th}^{2})^{2}}%
-2\sin(\widehat{\theta}_{+}+\phi)] \tag{48}%
\end{equation}
where the term of the time derivative is dropped for we are considering the
static case. We can change the variable of integration and get%

\begin{equation}
\mathcal{E}=\sqrt{2}\int_{\widehat{\theta}_{+,1}}^{\widehat{\theta}_{+,2}%
}d\widehat{\theta}_{+}[1+\frac{8A_{th}^{2}}{(1-A_{th}^{2})^{2}}-\sin
(\widehat{\theta}_{+}+\phi)]^{1/2}-L[1+\frac{8A_{th}^{2}}{(1-A_{th}^{2})^{2}}]
\tag{49}%
\end{equation}
where $\widehat{\theta}_{+,1}=\pi/2-\varphi+4\arctan(A_{th}sc(-\beta z_{0}))$
and $\widehat{\theta}_{+,2}=\pi/2-\varphi+4\arctan(A_{th}sc(\beta L-\beta
z_{0}))$. It has been shown in Appendix B that $\widehat{\theta}_{+,2}-$
$\widehat{\theta}_{+,1}=2\pi$. Therefore the total energy $\mathcal{E}$ is
independent of $z_{0}$ and $\phi$ because the integration is over an entire
period. $\mathcal{E}$ depends on only one parameter, $\beta$, for static
soliton because $A=A_{th}$\ and $A_{th}$\ is also determined by $\beta$.\ The
spectrum is plotted in Fig. 2 with eq. (45). It is very similar to that of
standing wave with $\beta$\ being the wave vector.

In the limit $L\rightarrow\infty$, $A_{th}$ becomes vanishingly small as it
can be seen from eqs. (44) and (41). We thus have $\beta$, $k_{f}\rightarrow1$
and $K(k_{f})\simeq\ln(4/A_{th}^{2})$. With eq. (41) we get
\begin{equation}
A_{th}\simeq2\exp(-L/2). \tag{50}%
\end{equation}
The magnitude of $A_{th}$\ is small even for a modest length of $L$.\ For
example,when $L=24$, $A_{th}\simeq2/e^{12}\simeq1.23\times10^{-5}$. Since
$A_{th}$\ can be viewed as the amplitude of the nonlinear wave,\ the wave form
on a long spin chain becomes flat everywhere except for narrow regions
$cn(z-z_{0})\sim0$.\ Therefore, one can expect a sudden change of
$\widehat{\theta}_{+}$ or a sharp domain wall.

\paragraph{5. Results and discussion}

In this section, we present the results of our calculation above. First of
all, we plotted $z_{0}$ versus $\phi$ in Fig. 3. $z_{0}$\ can be viewed as a
reference point of the solution of the sine-Gordon equation. Its movement is a
clear indication that the solitons is set in motion by $\phi$. Its motion is
not smooth as one can easily see that there is an abrupt change near $\phi
=\pi/2$, a manifestation of the nonlinearity of the solution. It decreases by
a distance $L$\ when $\phi$\ increases by $2\pi$. To see more clearly how the
soliton moves, we plotted in Figs. 4 $\widehat{\theta}_{+}$\ versus lattice
sites for different values of $\phi$ at $L=24$. We can go back to the original
spin system to see how spins are transported by utilizing eqs. (11) and (12).
Hence, $S_{j}^{z}$ and $S_{j+1}^{+}S_{j}^{-}+S_{j+1}^{-}S_{j}^{+}$ are also plotted.

In view of eq. (11), the \textquotedblright domain wall\textquotedblright\ or
the region where there is a jump of $\widehat{\theta}_{+}$\ is the place where
$\left\langle S^{z}\right\rangle $\ is large. Hence, in Figs. 4\ the jump of
$\theta_{+}$ and the peak of $\left\langle S^{z}\right\rangle $\ move together
with varying $\phi$ and spins moves from right to left. These figures also
show that the static soliton solution really is an N\'{e}el state in the spin
chain except in the neighborhood of $\phi\simeq\pi/2$. In this range, the
N\'{e}el state becomes unstable due to the dimer coupling $\Delta(t)$\ we
added.\ This is manifest in Figs. 4(b-d) where $S_{j+1}^{+}S_{j}^{-}%
+S_{j+1}^{1}S_{j}^{+}$ which is proportional to the dimer state amplitude is
large. Recall that $\phi$\ is defined in $(h_{st},\Delta)=R(\cos\phi,\sin
\phi)$. The dimer strength $\Delta\ $is the the largest when $\phi\simeq\pi
/2$. It is when the N\'{e}el state becomes unstable and the dimer state
amplitude\ becomes significant that the transport of spin becomes possible.
Not coincidentally, one can find in Fig. 3 that $z_{0}$\ changes abruptly in
this range.

In Fig. 5, $\widehat{\theta}_{+}$, $S_{j}^{z}$ and $S_{j+1}^{+}S_{j}%
^{-}+S_{j+1}^{-}S_{j}^{+}$\ versus lattice sites for $\phi=\pi/2$ at a shorter
length of $L$\ ($L=14$)\ is plotted for comparison.\ For smaller $L$, the
curve of $\widehat{\theta}_{+}$\ is smoother or the domain wall is not as
sharp. On the other hand, the edge (end) effect is more important. The
directions of spins are less ordered for a shorter spin chain because the edge
effect penetrates deeper into the "bulk".

We will elaborate more on how the spins are transported. This can be done with
eq. (11). The spin polarization is
\begin{equation}
P_{S^{z}}=\frac{1}{L}%
%TCIMACRO{\dint \limits_{0}^{L}}%
%BeginExpansion
{\displaystyle\int\limits_{0}^{L}}
%EndExpansion
zS^{z}(z)dz. \tag{51}%
\end{equation}
By integration by parts, we found
\begin{equation}
P_{S^{z}}=%
%TCIMACRO{\dint \limits_{0}^{L}}%
%BeginExpansion
{\displaystyle\int\limits_{0}^{L}}
%EndExpansion
S^{z}(z^{\prime})dz^{\prime}-\frac{1}{L}%
%TCIMACRO{\dint \limits_{0}^{L}}%
%BeginExpansion
{\displaystyle\int\limits_{0}^{L}}
%EndExpansion%
%TCIMACRO{\dint \limits_{0}^{z}}%
%BeginExpansion
{\displaystyle\int\limits_{0}^{z}}
%EndExpansion
S^{z}(z^{\prime})dz^{\prime}dz \tag{52}%
\end{equation}
To find the variation of $P_{S^{z}}$\ due to $\phi$\ we note that the first
term remain constant as $\phi$\ varies. This can be seen by substituting eq.
(11) into the integration. The contribution of the oscillatory term vanishes
as $L\rightarrow\infty$ and the term of the derivative gives unity due to our
boundary condition, no matter what the value of $\phi$\ is. Thus, denoting the
variation of $P_{S^{z}}$\ due to the adiabatic change of $\phi$ by $\delta
P_{S^{z}}$, we have
\begin{align}
\delta P_{S^{z}}  &  =-\frac{1}{L}\{[%
%TCIMACRO{\dint \limits_{0}^{L}}%
%BeginExpansion
{\displaystyle\int\limits_{0}^{L}}
%EndExpansion%
%TCIMACRO{\dint \limits_{0}^{z}}%
%BeginExpansion
{\displaystyle\int\limits_{0}^{z}}
%EndExpansion
S^{z}(z^{\prime})dz^{\prime}dz]|_{\phi=\phi_{2}}-[%
%TCIMACRO{\dint \limits_{0}^{L}}%
%BeginExpansion
{\displaystyle\int\limits_{0}^{L}}
%EndExpansion%
%TCIMACRO{\dint \limits_{0}^{z}}%
%BeginExpansion
{\displaystyle\int\limits_{0}^{z}}
%EndExpansion
S^{z}(z^{\prime})dz^{\prime}dz]|_{\phi=\phi_{1}}\}\nonumber\\
&  \simeq-\frac{1}{2\pi L}\{[%
%TCIMACRO{\dint \limits_{0}^{L}}%
%BeginExpansion
{\displaystyle\int\limits_{0}^{L}}
%EndExpansion
\widehat{\theta}_{+}(z)dz]|_{\phi_{2}}-[%
%TCIMACRO{\dint \limits_{0}^{L}}%
%BeginExpansion
{\displaystyle\int\limits_{0}^{L}}
%EndExpansion
\widehat{\theta}_{+}(z)dz]|_{\phi_{1}}\} \tag{53}%
\end{align}
\ \ where the second step can be reached by using the approximation
\begin{equation}
S^{z}\simeq(\partial\widehat{\theta}_{+}/\partial z)/2\pi\tag{54}%
\end{equation}
in integration for large $L$. In view of Figs. 4(b-d) where $\phi$ increases
exceeding $\pi/2$, we found that $\widehat{\theta}_{+}$\ increases\ in the
entire length of the system by approximately $2\pi$ and hence, $\delta
P_{S^{z}}\simeq-1$\ around $\varphi=\pi/2$ and a spin 1 is moved from right to
left around $\varphi=\pi/2$. In Figs. 4(e-h), where $\widehat{\theta}_{+}$ is
almost constant away from ends, we did not see any spin movement in the bulk
but rather, there are changes of spins at both ends.

To see the \textit{quantum} spin transport (a spin of unity being transported)
more clearly, we consider the limit $L\rightarrow\infty$ which can be simulate
very closely by the case $L=24$. The following equation will be very useful
for our purpose
\begin{equation}
\frac{\partial\widehat{\theta}_{+}}{\partial z}=\frac{4A_{th}\beta
dn(\beta(z-z_{0}))}{cn^{2}(\beta(z-z_{0}))+A_{th}^{2}sn^{2}(\beta(z-z_{0}))},
\tag{55}%
\end{equation}
since it is the dominant contribution to $S^{z}$. From eqs. (41), (50),
(A-11), (A-17) and (A-18), we find that
\begin{equation}
\frac{\partial\widehat{\theta}_{+}}{\partial z}\simeq\frac{8e^{-L/2}%
\cosh(z-z_{0})}{1+4e^{-L}\sinh^{2}(z-z_{0})} \tag{56}%
\end{equation}
The peak of $\partial\widehat{\theta}_{+}/\partial z$ or $S^{z}$\ is at
\begin{equation}
z-z_{0}\simeq L/2. \tag{57}%
\end{equation}
The larger $L$, the narrower the peak. On the other hand, $z_{0}$\ is
determined by the boundary condition. When $\phi=\pi/2-\delta$\ where
$\delta\ $is\ a small and\ positive\ number, we have
\begin{equation}
\tan(-\frac{\delta}{4})=A_{th}sc(-\beta z_{0}) \tag{58}%
\end{equation}
and we find that $z_{0}\simeq L/2$ as long as $\delta$\ remains finite (see
eq. (50)).\ The resulting $S^{z}$ due to eqs. (11) and (55) has a peak at the
right end and vanishes everywhere else.\ When $\phi=\pi/2$, eq. (57) gives
$z_{0}=0$ and the peak of $S^{z}$\ moves to the center of the spin chain.\ If
$\phi=\pi/2+\delta$,\ then $z_{0}\simeq-L/2$, and the peak moves to the left end.\ 

As for the quantity of spin transported, we can analyze the variation of
$\widehat{\theta}_{+}$.\ In view of eq. (40), as $\phi=\pi/2-\delta$,
$\widehat{\theta}_{+}$\ almost vanishes for the entire chain except for the
right end. As $\phi$ increases to $\pi/2+\delta$, $\widehat{\theta}_{+}%
\simeq2\pi$\ for the entire chain except for the left end where it drops to
zero sharply. Hence, according to eq. (53), during the interval $\phi
=\pi/2-\delta$ to $\phi=\pi/2+\delta$, a spin of \textit{unity} is transported
from the right end to the left end. As $\phi$ increases onward from\ $\pi
/2+\delta$,\ there is no spin transport in the bulk.\ Nevertheless, the
plateau of $\widehat{\theta}_{+}$\ is lowered (see Figs. 4(e-h)) and as a
result, $S^{z}$\ at the left end decreases and a peak of $S^{z}$\ at the right
end start to grow. This kind of change\ continues until $\phi$\ reaches
$5\pi/2-\delta$.\ At this stage the state of soliton returns to that of
$\phi=\pi/2-\delta$. We conclude the analysis of our result by the following
summary: There is a swift spin transport in the bulk during the short interval
between $\phi=\pi/2-\delta$\ and $\phi=\pi/2+\delta$ where $\delta$\ can be
made arbitrarily small if $L\rightarrow\infty$. The net spin transported is
unity.\ Beyond this interval, the spins at both ends vary with $\phi$\ but
there is no spin change in the bulk.

The question will inevitably be raised: Can spin be transported? We have seen
that the soliton returns to the starting state if $\phi$\ increases by $2\pi
$.\ Thus there is no net spin transported in a cycle. However, in a realistic
system, two ends of the spin chain must be connected to leads. The leads ought
to serve as a spin source and a spin drain. Thus it is reasonable to envisage
the following picture: At $\phi=\pi/2+\delta$\ the left end can dump spin into
a spin drain and the right end can extract spin from the source.\ When
$\phi=5\pi/2-\delta$\ which is equivalent to $\phi=\pi/2-\delta$,\ the dumping
of spin at the left end is complete and the peak of spin at the right end has
grown into saturation. Then an unity of spin is transported from the right end
to the left end when $\phi$\ increases from $\phi=\pi/2-\delta$\ to\ $\phi
=\pi/2+\delta$. This is in all intent and purpose, same as a physical system
of spin transport.\ On this point, our system is same as the $Z_{2}$\ spin
pump proposed by Fu and Kane[16]. However, there is an important difference.
Our spin transport is through a bulk state. This is completely different from
Shindou's[14] and Fu and Kane's[15,16]\ pictures in which the level crossing
of the end (edge) states is essential. Consequently, our spin transport is
quantized because it is through a bulk state. Connecting to spin reservoir
cannot destroy the quantization as it will do the transport due to end states.\ 

This work is supported in part by NSC of Taiwan, ROC under the contract number
NSC 95-2112-M-002-048-MY3.

\paragraph{Appendix A:}

In this appendix we listed some properties of the Jacobian elliptic functions.
See eqs. (24-25) for the definition.
\begin{align}
sn^{2}(u)+cn^{2}(u)  &  =1,\tag{A-1}\\
dn^{2}(u)+k^{2}sn^{2}(u)  &  =1,\tag{A-2}\\
sn(u+K)  &  =cn(u)/dn(u)\tag{A-3}\\
cn(u+K)  &  =-sn(u)/dn(u) \tag{A-4}%
\end{align}
where $k^{\prime2}=1-k^{2}$. The derivatives of Jacobian elliptic functions
are
\begin{align}
\partial_{u}sn(u)  &  =cn(u)dn(u),\tag{A-5}\\
\partial_{u}cn(u)  &  =-sn(u)dn(u),\tag{A-6}\\
\partial_{u}dn(u)  &  =-k^{2}sn(u)cn(u). \tag{A-7}%
\end{align}
Using above equations we found the differential equations to be satisfied by
the Jacobian elliptic functions and listed them in Table 1 where
\begin{align}
ns(u)  &  =1/sn(u),\tag{A-8}\\
nc(u)  &  =1/cn(u),\tag{A-9}\\
nd(u)  &  =1/dn(u),\tag{A-10}\\
sc(u)  &  =sn(u)/cn(u),\tag{A-11}\\
sd(u)  &  =sn(u)/dn(u),\tag{A-12}\\
cd(u)  &  =cn(u)/dn(u),\tag{A-13}\\
cs(u)  &  =1/sc(u),\tag{A-14}\\
ds(u)  &  =1/sd(u),\tag{A-15}\\
dc(u)  &  =1/cd(u). \tag{A-16}%
\end{align}
Having checked those equations in Table I, one can see that there are many
combinations of JEF that can satisfy eqs. (20) and (21) where the discriminant
is$\ \mu^{2}+4\kappa\lambda$\ for $k_{f}$\ and $(\mu+1)^{2}+4\kappa\lambda
$\ for $k_{g}$.\ 

For $L\longrightarrow\infty$, we find from eq. (44) that $k\rightarrow1$ and%
\begin{equation}
sn(u)\simeq\tanh(u), \tag{A-17}%
\end{equation}%
\begin{equation}
cn(u)\simeq dn(u)\simeq\sec h(u), \tag{A-18}%
\end{equation}

\paragraph{Appendix B}

In this Appendix, we prove the following lemma.

\begin{lemma}
If $\theta_{+}(z)=\pi/2+\varphi+4\arctan\{A_{th}sc[\beta(L-z_{0});k_{f}]\}$
and $A=A_{th}$, then $\widehat{\theta}_{+}(z=L)-\widehat{\theta}%
_{+}(z=0)=2l\pi$ where\ $l$ is a natural\ number defined by
\begin{equation}
\beta L=lK(k_{f}=\sqrt{1-A_{th}^{4}})=n\int_{0}^{1}\frac{dt}{\sqrt
{(1-t^{2})[1-(1-A_{th}^{4})t^{2}]}} \tag{B-1}%
\end{equation}
with $\beta=1/(1-A_{th}^{2})$.
\end{lemma}

\noindent Proof:

The boundary condition requires that at $z=0$
\begin{equation}
\frac{\phi}{4}=-\arctan A_{th}sn(-\beta z_{0}) \tag{B-2}%
\end{equation}
and at $z=L$\
\begin{equation}
\frac{n\pi}{2}-\frac{\phi}{4}=\arctan A_{th}sc(\beta(L-z_{0}))=\arctan
A_{th}sc(lK-\beta z_{0}) \tag{B-3}%
\end{equation}
Note that the period of $sc(u)$ is $2K$. Hence, when $z$\ increases from
$0$\ to $L$,\ $l/2$\ periods will pass.\ Since the period of $\arctan
$\ function is $\pi$, the term $l\pi/2$\ has to be added on the left hand side
of eq. (C-3). To see it more explicitly, define
\begin{equation}
\alpha=\arctan(A_{th}sc(\beta(L-z_{0})))=\arctan(A_{th}sc(lK-\beta z_{0})),
\tag{B-4}%
\end{equation}
and
\begin{equation}
\varsigma=\arctan(A_{th}sc(-\beta z_{0})), \tag{B-5}%
\end{equation}
then we have
\begin{equation}
\tan(\alpha-\varsigma)=\frac{\tan\alpha-\tan\varsigma}{1+\tan\alpha
\tan\varsigma}=A_{th}\frac{sc(lK-\beta z_{0})-sc(-\beta z_{0})}{1+A_{th}%
^{2}sc(lK-\beta z_{0})sc(-\beta z_{0})}. \tag{B-6}%
\end{equation}
With eqs. (A-3) and (A-4), we have
\begin{equation}
sc(\beta(nK-z_{0}))=\frac{sn(nK-\beta z_{0})}{cn(nK-\beta z_{0})}%
=\frac{cn(nK-K-\beta z_{0}))}{-k_{f}^{\prime}sn(nK-K-\beta z_{0}))}. \tag{B-7}%
\end{equation}
where $k_{f}^{2}+k_{f}^{\prime2}=1$. Since $k_{f}^{2}=1-A_{th}^{4}$,\ we find
that $k_{f}^{\prime}=A_{th}^{2}$.\ Eq. (C-6) becomes
\begin{equation}
\tan(\alpha-\varsigma)=-A_{th}\frac{cs(lK-K-\beta z_{0})/k_{f}^{\prime
}-sc(-\beta z_{0})}{1-sc(-\beta z_{0})cs(lK-K-\beta z_{0})} \tag{B-8}%
\end{equation}
where $cs(u)=1/sc(u)$. So the denominator of $\tan(\alpha-\varsigma)$ vanishes
as $n=1$ and the numerator is finite. Thus $\tan(\alpha-\varsigma)=\pm\infty$
and $\alpha-\varsigma=\pm\pi/2$. The sign is determined by the fact that as
$z$\ increases, $\widehat{\theta}_{+}$ also increases, thus the positive sign
should be chosen and $\widehat{\theta}_{+}(z=L)-\widehat{\theta}_{+}%
(z=0)=2\pi$. If $l=2$,\ we can use eqs. (A-3) and (A-4) again or simply use
the fact that the period of $sc(u)$\ is $2K$ to find out that the numerator
vanishes. Thus, $\tan(\alpha-\varsigma)=\pi$ and $\widehat{\theta}%
_{+}(z=L)-\widehat{\theta}_{+}(z=0)=4\pi$. Hence, we conclude that
$\widehat{\theta}_{+}(z=L)-\widehat{\theta}_{+}(z=0)=4(\alpha-\varsigma
)=2l\pi.$

\noindent End of Proof.

\paragraph{References:}

[1]. D. J. Thouless, Phy. Rev. \textbf{B27}, 6083 (1983).

[2]. Q. Niu and D. J. Thouless, J. Phys. A, \textbf{17}, 2453 (1984).

[3]. P. W. Brouwer, Phys. Rev. \textbf{B58}, 10135 (1998).

[4]. T. A. Shutenko, I. L. Aleiner and B. L. Altshuler, Phys. Rev.
\textbf{B61}, 10366 (2000).

[5]. Y. Levinson, Entin-O.Wohlman and P. Wolfe, Physica \textbf{A302}, 335
(2001); Wohlman-O.Entin and A. Aharony, Phys. Rev. \textbf{B66}, 35329 (2002).

[6]. M. Blaauboer, Phys. Rev. \textbf{B65}, 235318 (2002).

[7]. J. Wang and B. Wang, Phys. Rev. \textbf{B65}, 153311 (2002); B. Wang and
J. Wang, Phys. Rev. \textbf{B66}, 201305 (2002).

[8]. P. Sharma and C. Chamon, Phys. Rev. Lett. \textbf{87}, 96401 (2001) and cond-mat/0209291.

[9]. R. Citro, N. Anderi and Q. Niu, cond-mat/0306181.

[10]. D. J. Thouless, M. Kohmoto, M. P. Nightingale and M. den Nijs, Phys.
Rev. Lett. \textbf{49}, 405 (1982).

[11]. B. I. Halperin, Phys. Rev. \textbf{B25}, 2185 (1982).

[12]. Y. Hatsugai, Phys. Rev. Lett. \textbf{71}, 3697 (1993)

[13]. Y. Hatsugai, Phys. Rev. \textbf{B48}, 1185 (1993).

[14]. R. Shindou, J. Phys. Soc.Jpn. \textbf{74}, 1214 (2005).

[15]. C. L. Kane and E. J. Mele , Phys. Rev. Lett. 95, 146802 (2005).

[16]. L. Fu and C. L. Kane, Phys. Rev. \textbf{B74}, 195312 (2006).

[17]. E. Fradkin, 'Field Theories of Condensed Matter Systems',
Addison-Wesley, Redwood City, CA (1991).

[18]. N. Nagaosa, 'Quantum Field Theory in Stongly Correlated Electronic
Systems', Springer-Verlag, Berlin (1999).

[19]. A. O. Gogolin, A. A. Nersesyan and A. M. Tsvelik, 'Bosonization and
Strongly Correlated Systems', Cambridge University Press (1998).

[20]. T. Giamarchi, 'Quantum Physics in One Dimension', Oxford University
Press (2004).

[21]. G. Costabile et. al., App. Phys. Lett. \textbf{32}, 587 (1978).

[22]. R. M. DeLeonardis et. al., J. App. Phys. \textbf{51}, 1211(1982).

[23]. R. M. DeLeonardis et. al., J. App. phys. \textbf{53}, 699 (1982).

[24]. Derek. F. Lawden, 'Elliptic Function and Application', Springer-Verlag
NY (1989).

\newpage

Fig captions:

Fig. 1 Threshold amplitude $(A_{th})$ versus system lengrh $L$ for the static
soliton solution.

Fig. 2 Energy spectrum of the static soliton versus $n=\beta/L$\ with the
boundary condition $\theta_{+}(z=0)=0$ and $\theta_{+}(z=L)=2n\pi$ when $L=24$.

Fig. 3 $z_{0}$ versus adiabatical paramter $\varphi$ with $L=24$ for static
soliton and boundary conditions $\theta_{+}(z=0)=0$ and $\theta_{+}%
(z=L)=2\pi.$

Fig. 4 $\widehat{\theta}_{+}$, $S_{z}$ and dimer state amplitude $S^{+}%
S^{-}+S^{-}S^{+}$ versus lattice sites with $L=24$ and different values of
$\phi$\ (a) $\phi=0$, (b) $\phi=\pi/4$, (c) $\phi=\pi/2$, (d) $\phi=3\pi/4$,
(e) $\phi=\pi$, (f) $\phi=5\pi/4$, (g) $\phi=3\pi/2$, (h) $\phi=3\pi/2$.

Fig. 5 $\widehat{\theta}_{+}$, $S_{z}$ and dimer state amplitude $S^{+}%
S^{-}+S^{-}S^{+}$ versus lattice sites with $L=14$ and $\phi=\pi/2$.

\newpage

Fig. 1%

%TCIMACRO{\FRAME{ftbpF}{4.1148in}{5.7934in}{0in}{}{}{fig1ath.eps}%
%width 4.1148in;  height 5.7934in;  depth 0in;  original-width 7.9373in;
%original-height 11.1976in;  cropleft "0";  croptop "1";  cropright "1";
%cropbottom "0";  filename 'fig1ath.EPS';file-properties "XNPEU";}}}%
%BeginExpansion
\begin{figure}
[ptb]
\begin{center}
\includegraphics[
height=5.7934in,
width=4.1148in
]%
{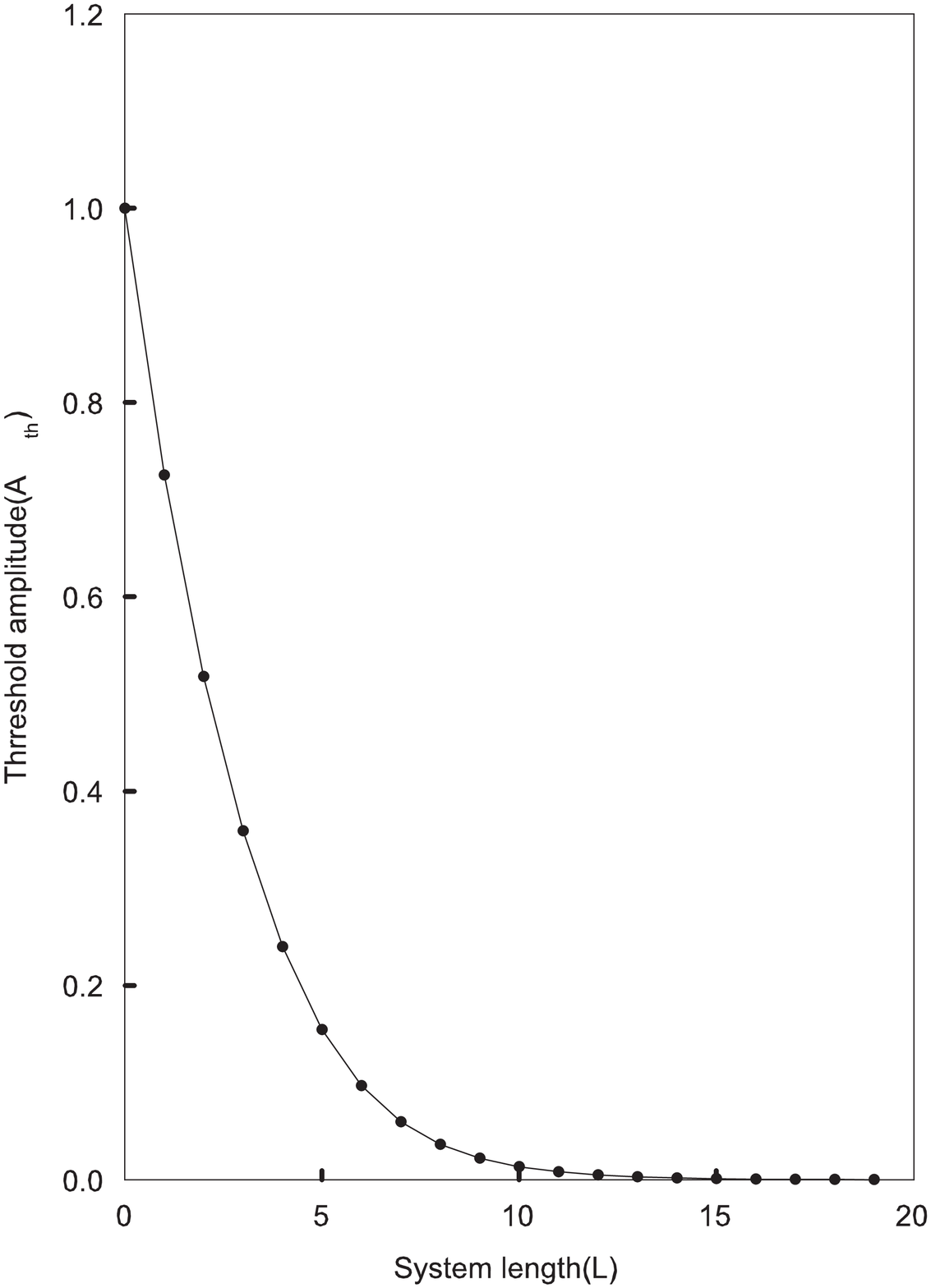}%
\end{center}
\end{figure}
%EndExpansion
\newpage

Fig. 2%

%TCIMACRO{\FRAME{ftbpF}{4.139in}{5.86in}{0pt}{}{}{fig2e.eps}%
%width 4.139in;  height 5.86in;  depth 0pt;  original-width 2.7337in;
%original-height 3.8821in;  cropleft "0";  croptop "1";  cropright "1";
%cropbottom "0";  filename 'fig2e.EPS';file-properties "XNPEU";}}}%
%BeginExpansion
\begin{figure}
[ptb]
\begin{center}
\includegraphics[
height=5.86in,
width=4.139in
]%
{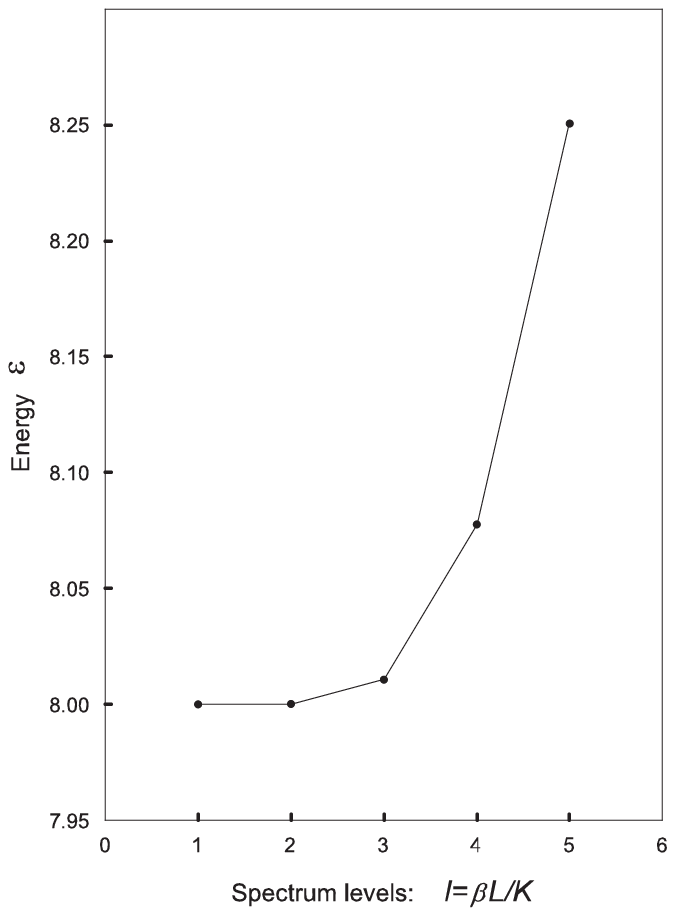}%
\end{center}
\end{figure}
%EndExpansion
\newpage

Fig. 3%

%TCIMACRO{\FRAME{ftbpF}{5.4613in}{3.8666in}{0pt}{}{}{fig3z0.eps}%
%width 5.4613in;  height 3.8666in;  depth 0pt;  original-width 3.8847in;
%original-height 2.7423in;  cropleft "0";  croptop "1";  cropright "1";
%cropbottom "0";  filename 'fig3z0.EPS';file-properties "XNPEU";}}}%
%BeginExpansion
\begin{figure}
[ptb]
\begin{center}
\includegraphics[
height=3.8666in,
width=5.4613in
]%
{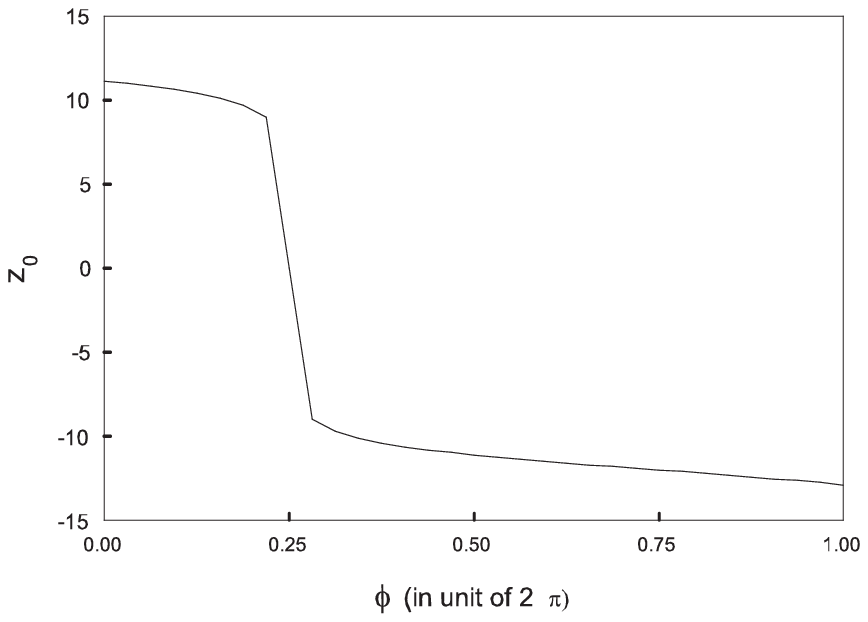}%
\end{center}
\end{figure}
%EndExpansion
\newpage

Fig. 4(a)%

%TCIMACRO{\FRAME{ftbpF}{4.3016in}{6.4307in}{0pt}{}{}{fig4aszfi0.eps}%
%width 4.3016in;  height 6.4307in;  depth 0pt;  original-width 7.8222in;
%original-height 11.7225in;  cropleft "0";  croptop "1";  cropright "1";
%cropbottom "0";  filename 'fig4aszfi0.EPS';file-properties "XNPEU";}}}%
%BeginExpansion
\begin{figure}
[ptb]
\begin{center}
\includegraphics[
height=6.4307in,
width=4.3016in
]%
{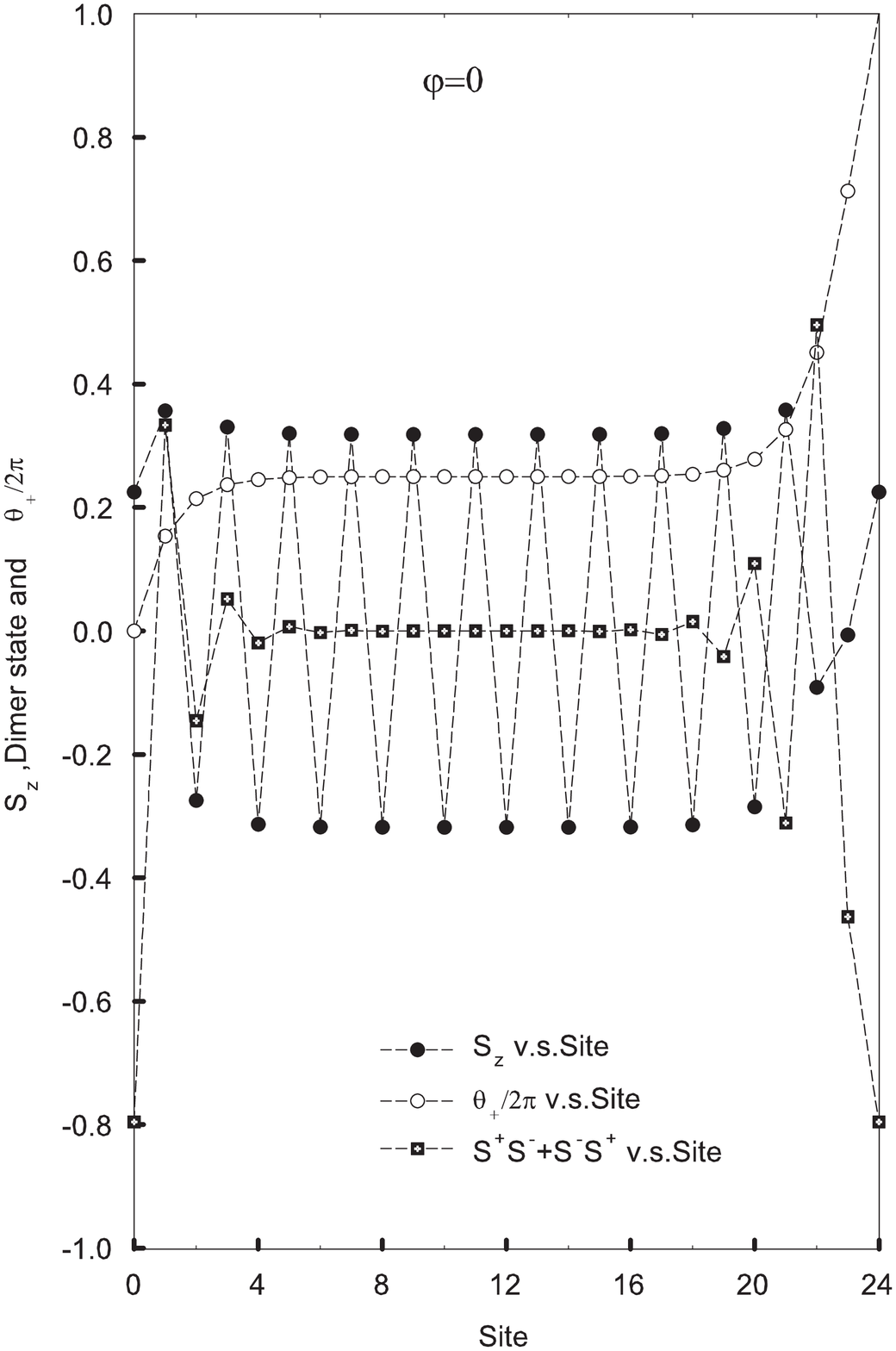}%
\end{center}
\end{figure}
%EndExpansion
\newpage

Fig. 4(b)%

%TCIMACRO{\FRAME{ftbpF}{4.2696in}{5.7917in}{0in}{}{}{fig4bszfi025.eps}%
%width 4.2696in;  height 5.7917in;  depth 0in;  original-width 8.2806in;
%original-height 11.2512in;  cropleft "0";  croptop "1";  cropright "1";
%cropbottom "0";  filename 'fig4bszfi025.EPS';file-properties "XNPEU";}}}%
%BeginExpansion
\begin{figure}
[ptb]
\begin{center}
\includegraphics[
height=5.7917in,
width=4.2696in
]%
{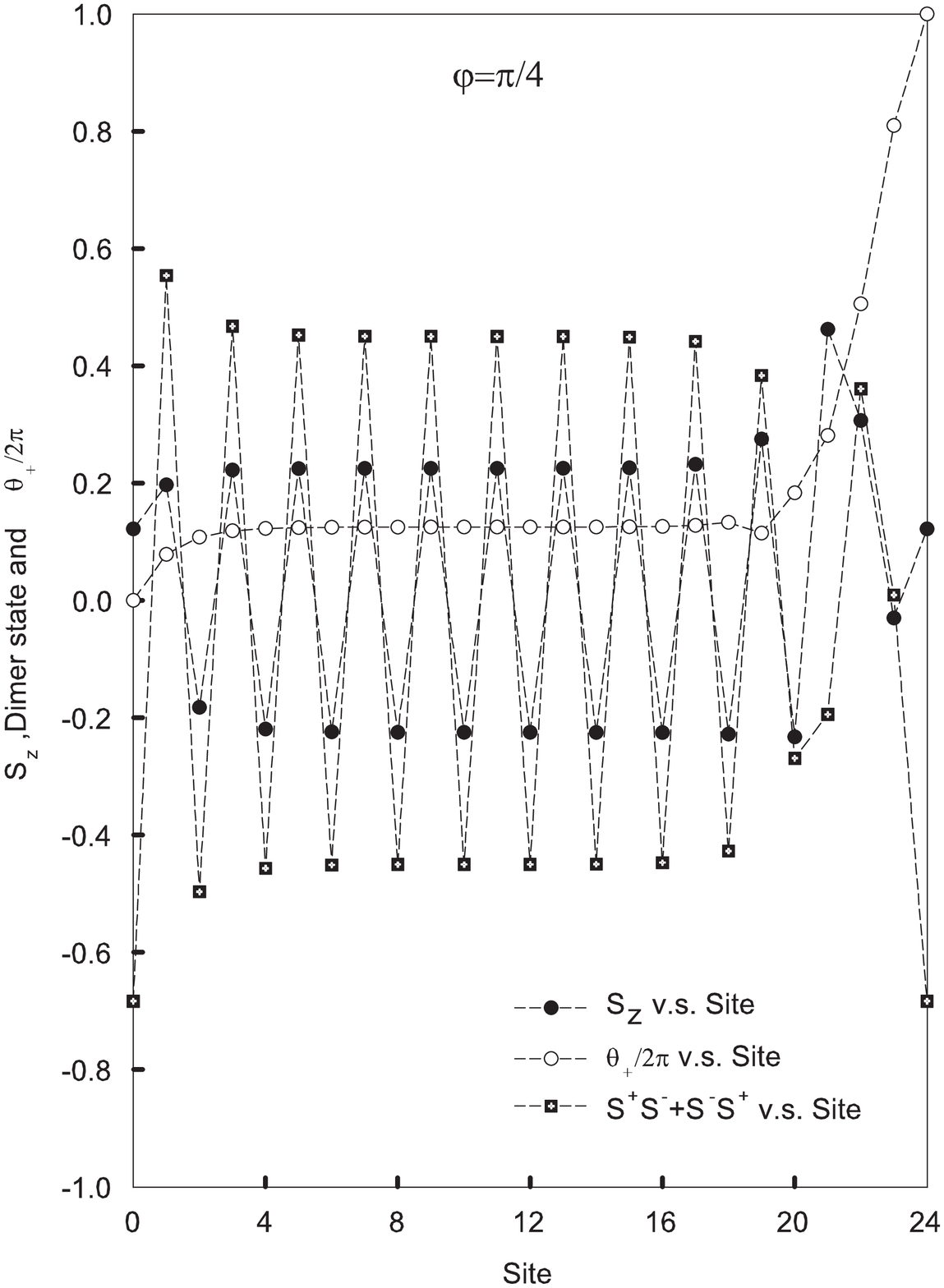}%
\end{center}
\end{figure}
%EndExpansion
\newpage

Fig. 4(c)%

%TCIMACRO{\FRAME{ftbpF}{4.1373in}{5.7934in}{0in}{}{}{fig4cszfi050.eps}%
%width 4.1373in;  height 5.7934in;  depth 0in;  original-width 8.406in;
%original-height 11.7917in;  cropleft "0";  croptop "1";  cropright "1";
%cropbottom "0";  filename 'fig4cszfi050.EPS';file-properties "XNPEU";}}}%
%BeginExpansion
\begin{figure}
[ptb]
\begin{center}
\includegraphics[
height=5.7934in,
width=4.1373in
]%
{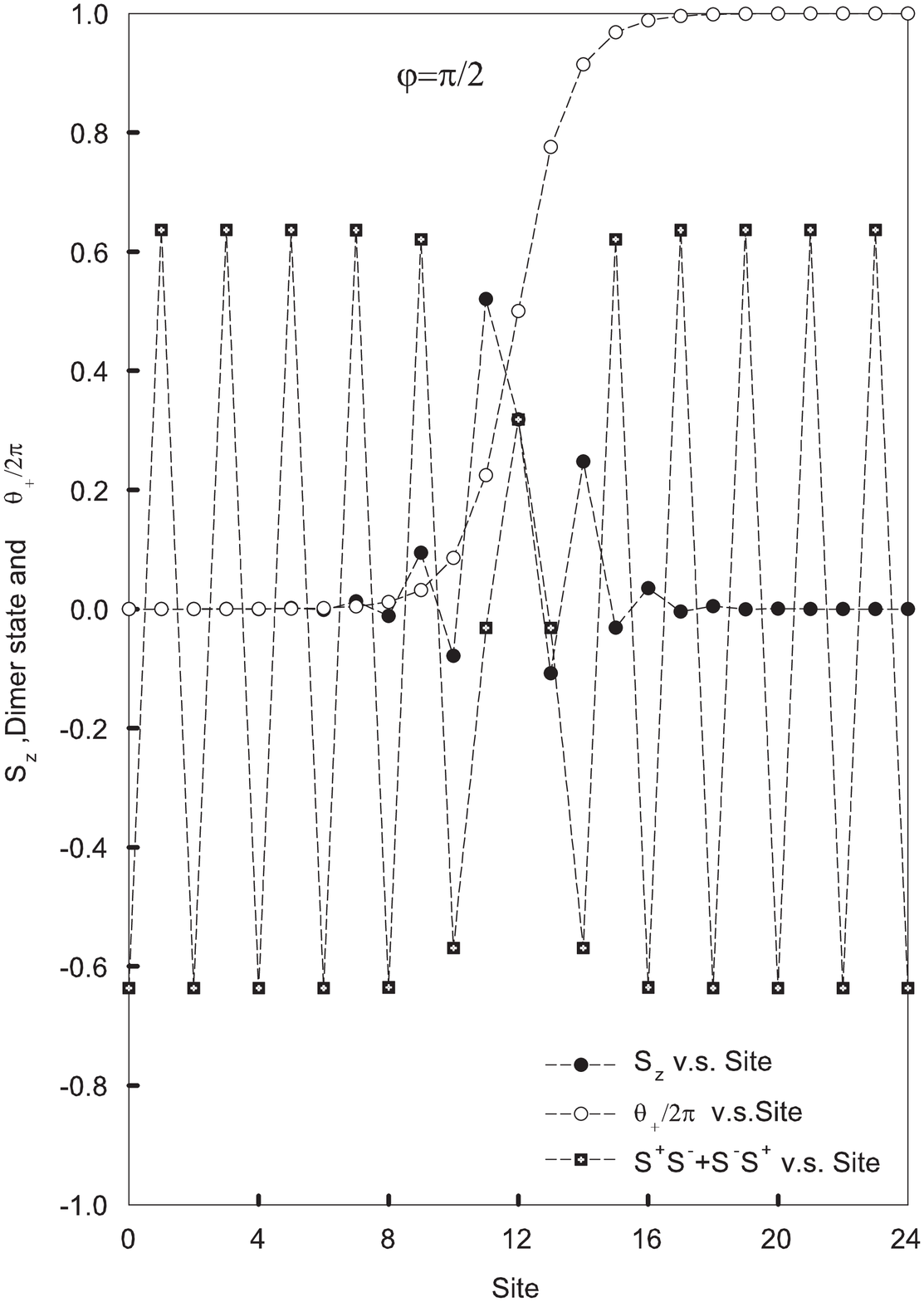}%
\end{center}
\end{figure}
%EndExpansion
\newpage

Fig. 4(d)%

%TCIMACRO{\FRAME{ftbpF}{4.1321in}{5.7943in}{0in}{}{}{fig4dszfi075.eps}%
%width 4.1321in;  height 5.7943in;  depth 0in;  original-width 8.323in;
%original-height 11.6949in;  cropleft "0";  croptop "1";  cropright "1";
%cropbottom "0";  filename 'fig4dszfi075.EPS';file-properties "XNPEU";}}}%
%BeginExpansion
\begin{figure}
[ptb]
\begin{center}
\includegraphics[
height=5.7943in,
width=4.1321in
]%
{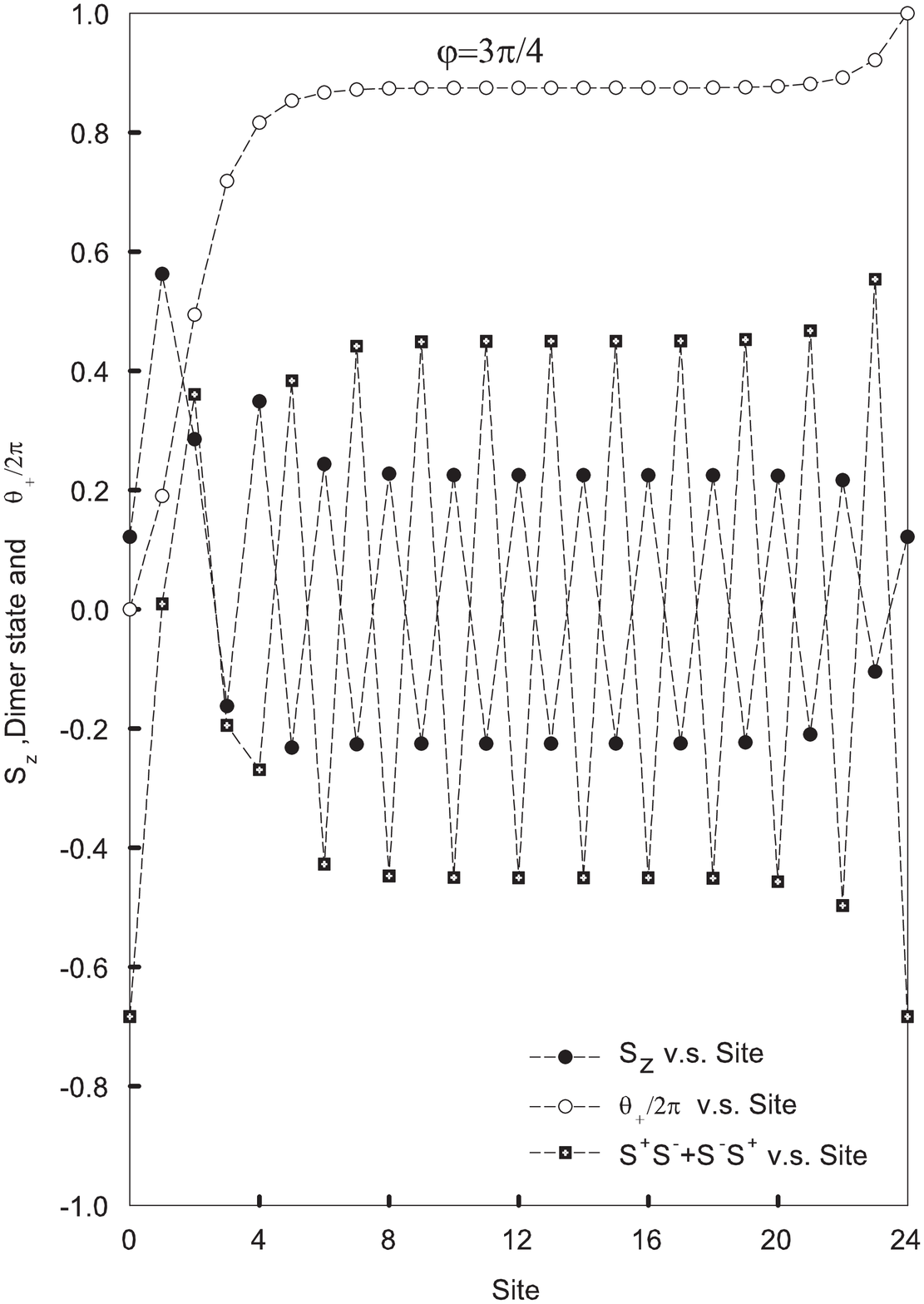}%
\end{center}
\end{figure}
%EndExpansion
\newpage

Fig. 4(e)%

%TCIMACRO{\FRAME{ftbpF}{4.094in}{5.7917in}{0in}{}{}{fig4eszfi100.eps}%
%width 4.094in;  height 5.7917in;  depth 0in;  original-width 8.2408in;
%original-height 11.681in;  cropleft "0";  croptop "1";  cropright "1";
%cropbottom "0";  filename 'fig4eszfi100.EPS';file-properties "XNPEU";}}}%
%BeginExpansion
\begin{figure}
[ptb]
\begin{center}
\includegraphics[
height=5.7917in,
width=4.094in
]%
{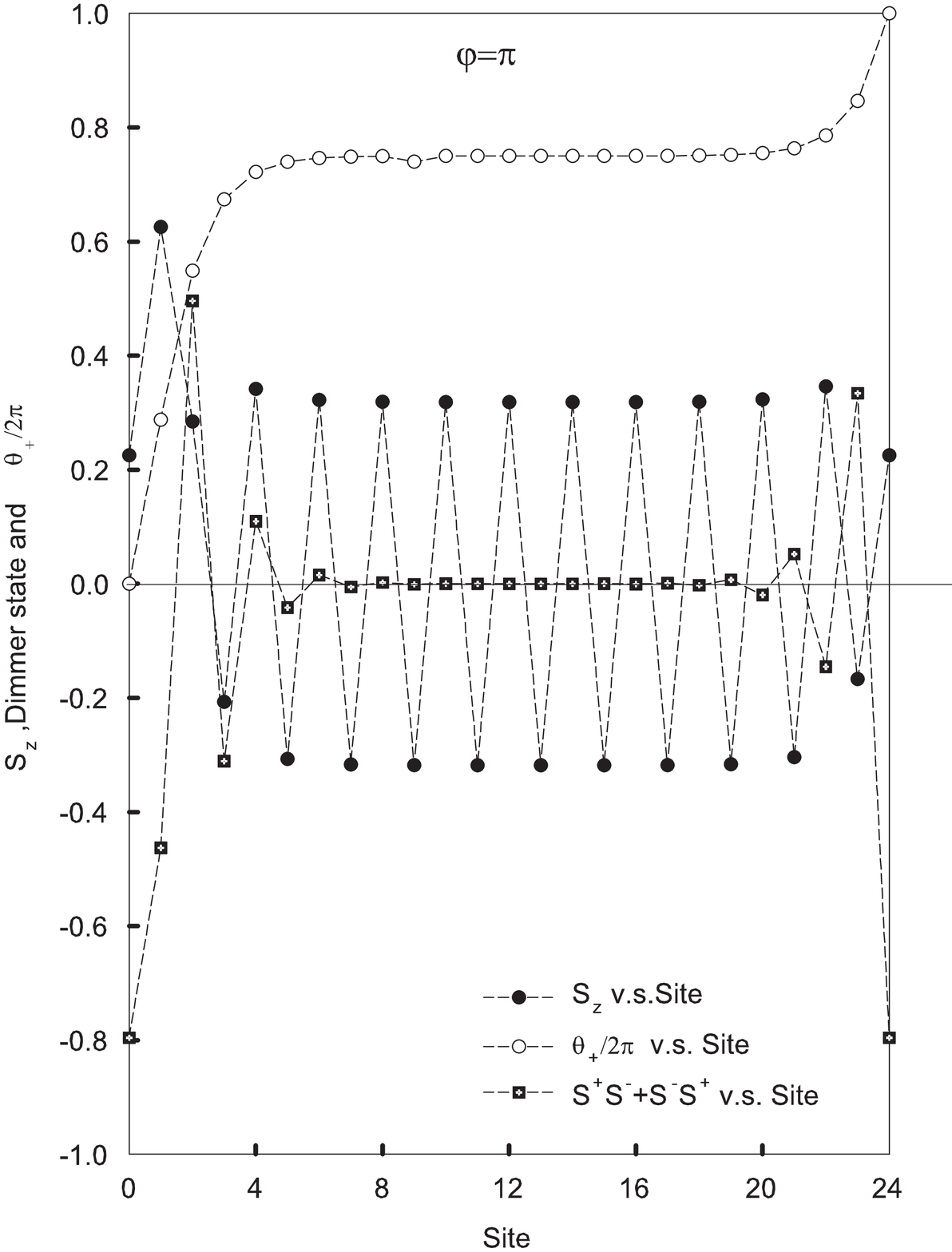}%
\end{center}
\end{figure}
%EndExpansion
\newpage

Fig. 4(f)%

%TCIMACRO{\FRAME{ftbpF}{4.171in}{5.7925in}{0in}{}{}{fig4fszfi125.eps}%
%width 4.171in;  height 5.7925in;  depth 0in;  original-width 8.3359in;
%original-height 11.5963in;  cropleft "0";  croptop "1";  cropright "1";
%cropbottom "0";  filename 'fig4fszfi125.EPS';file-properties "XNPEU";}}}%
%BeginExpansion
\begin{figure}
[ptb]
\begin{center}
\includegraphics[
height=5.7925in,
width=4.171in
]%
{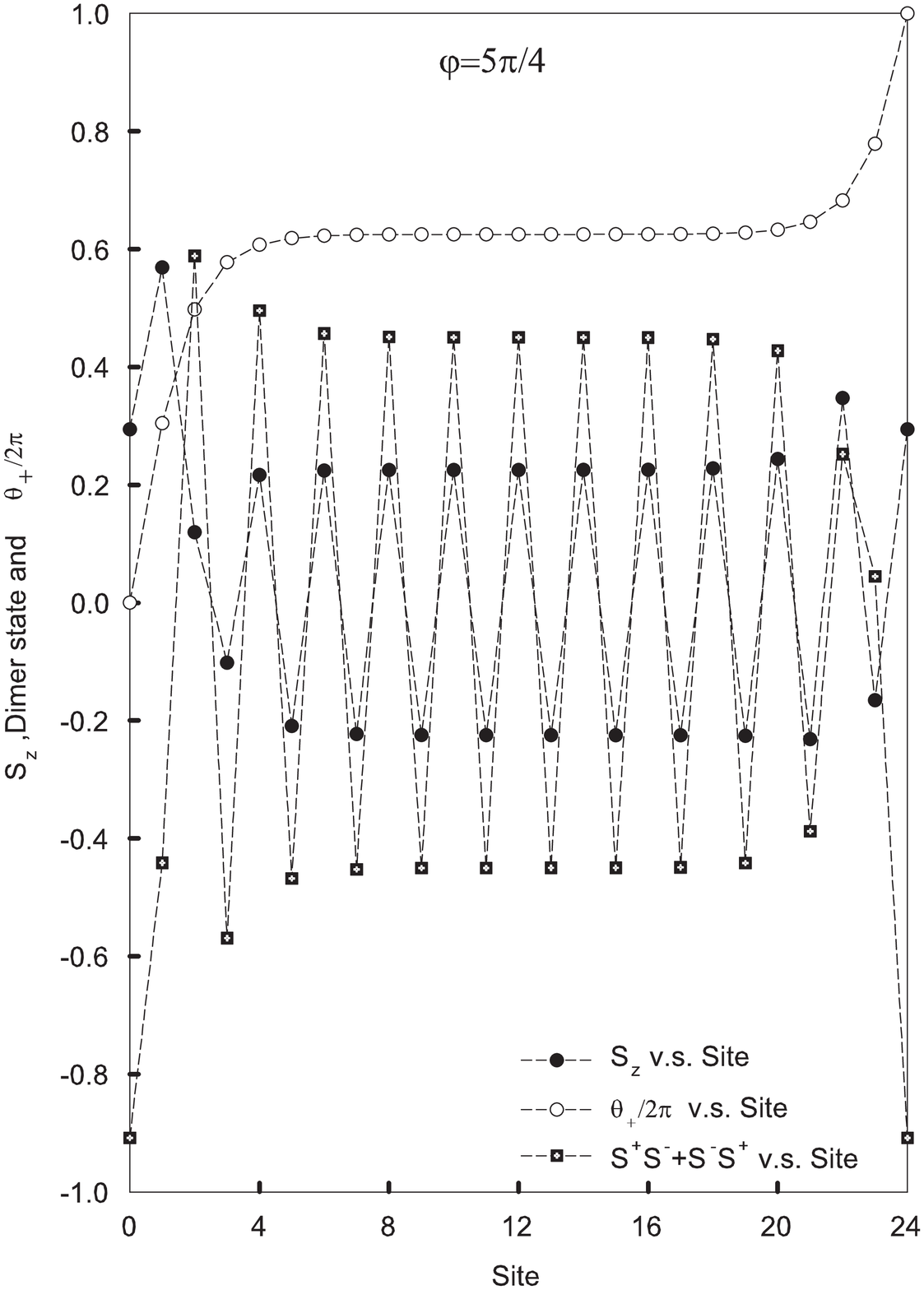}%
\end{center}
\end{figure}
%EndExpansion
\newpage

Fig. 4(g)%

%TCIMACRO{\FRAME{ftbpF}{4.1295in}{5.7917in}{0in}{}{}{fig4gszfi150.eps}%
%width 4.1295in;  height 5.7917in;  depth 0in;  original-width 8.1552in;
%original-height 11.4588in;  cropleft "0";  croptop "1";  cropright "1";
%cropbottom "0";  filename 'fig4gszfi150.EPS';file-properties "XNPEU";}}}%
%BeginExpansion
\begin{figure}
[ptb]
\begin{center}
\includegraphics[
height=5.7917in,
width=4.1295in
]%
{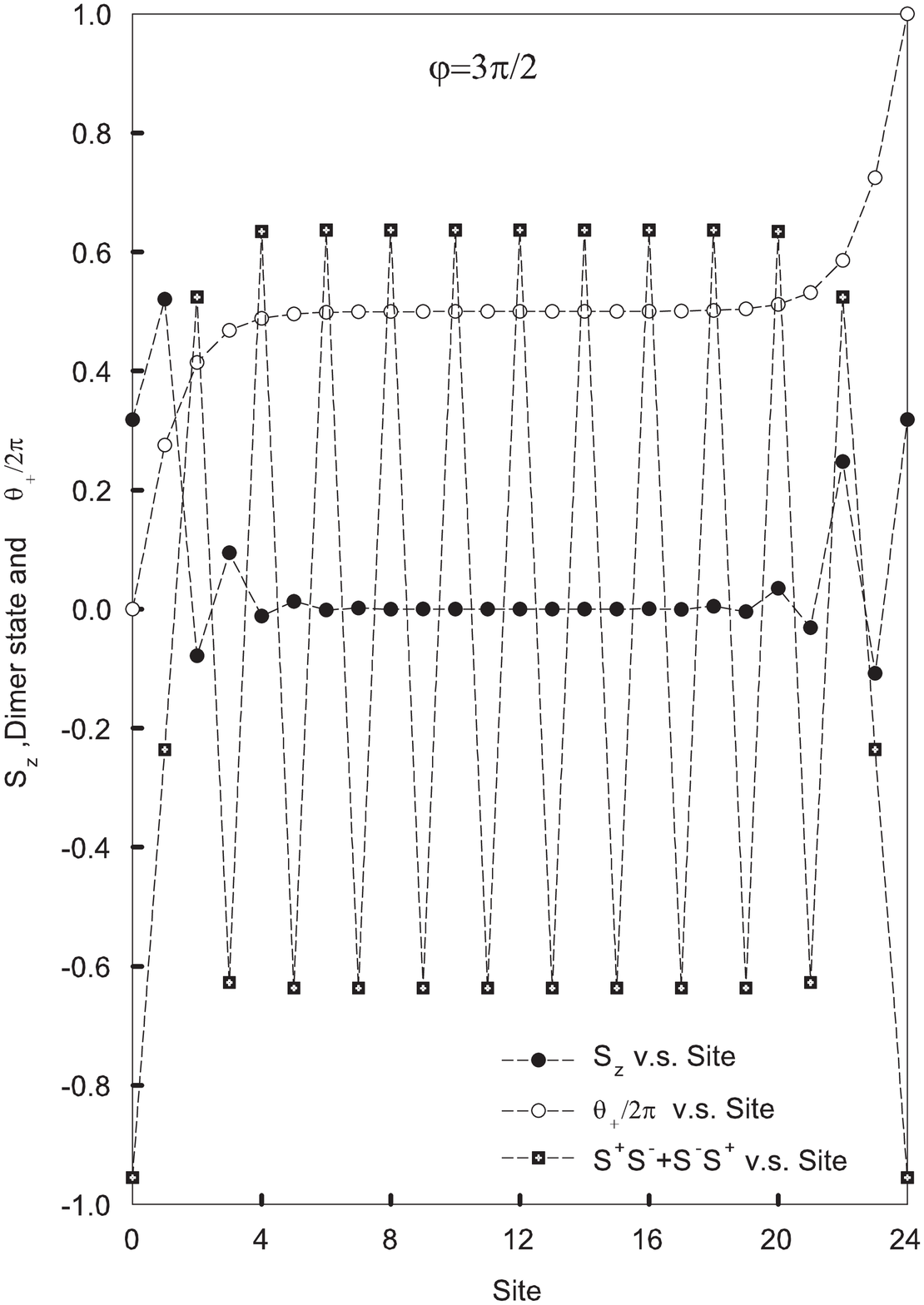}%
\end{center}
\end{figure}
%EndExpansion
\newpage

Fig. 4(h)%

%TCIMACRO{\FRAME{ftbpF}{4.1719in}{5.7934in}{0in}{}{}{fig4hszfi175.eps}%
%width 4.1719in;  height 5.7934in;  depth 0in;  original-width 8.2961in;
%original-height 11.5418in;  cropleft "0";  croptop "1";  cropright "1";
%cropbottom "0";  filename 'fig4hszfi175.EPS';file-properties "XNPEU";}}}%
%BeginExpansion
\begin{figure}
[ptb]
\begin{center}
\includegraphics[
height=5.7934in,
width=4.1719in
]%
{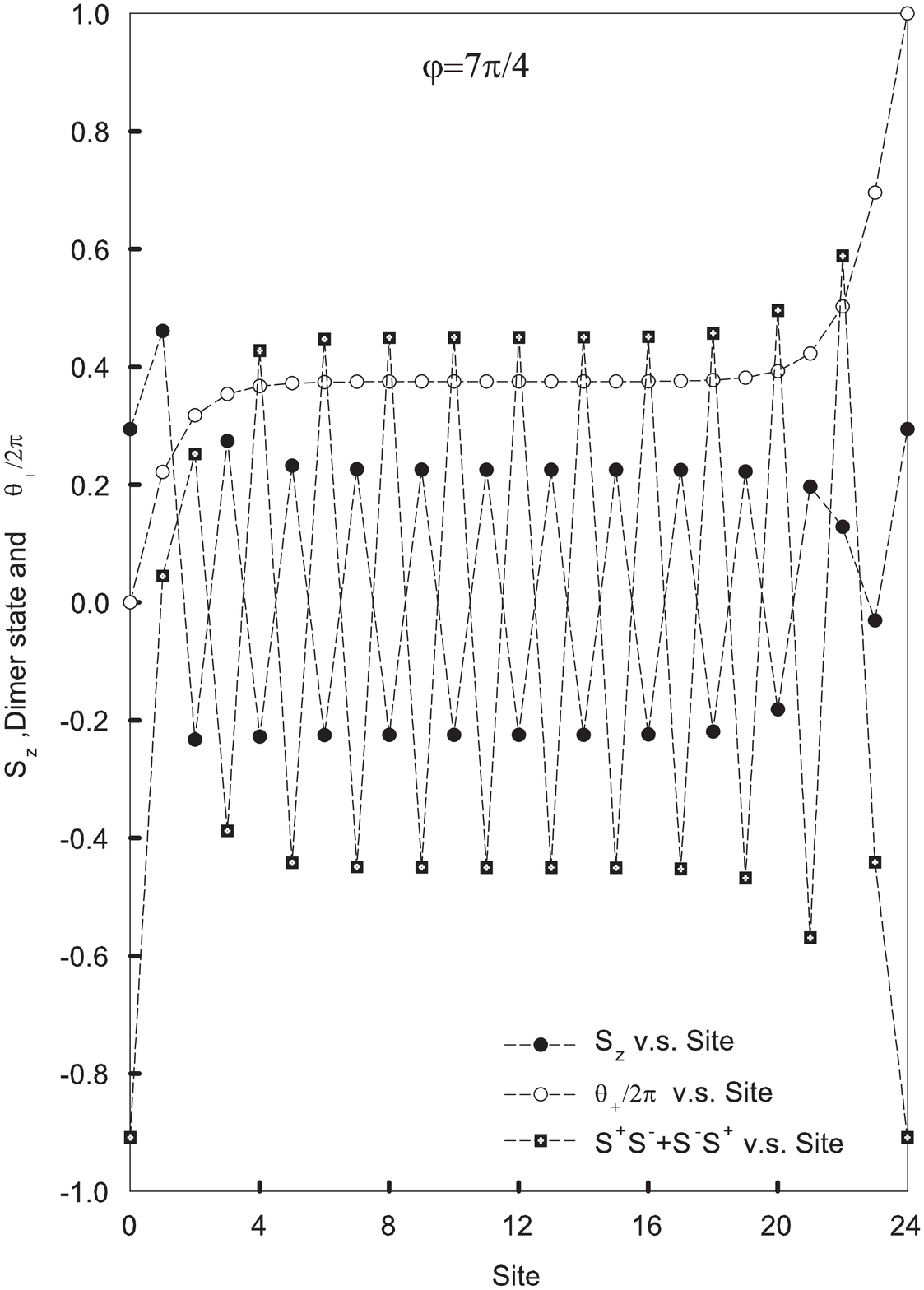}%
\end{center}
\end{figure}
%EndExpansion
\newpage

Fig. 5 \ %

%TCIMACRO{\FRAME{ftbpF}{4.241in}{5.7943in}{0in}{}{}{fig5szfi050.eps}%
%width 4.241in;  height 5.7943in;  depth 0in;  original-width 8.2408in;
%original-height 11.278in;  cropleft "0";  croptop "1";  cropright "1";
%cropbottom "0";  filename 'fig5szfi050.EPS';file-properties "XNPEU";}}}%
%BeginExpansion
\begin{figure}
[ptb]
\begin{center}
\includegraphics[
height=5.7943in,
width=4.241in
]%
{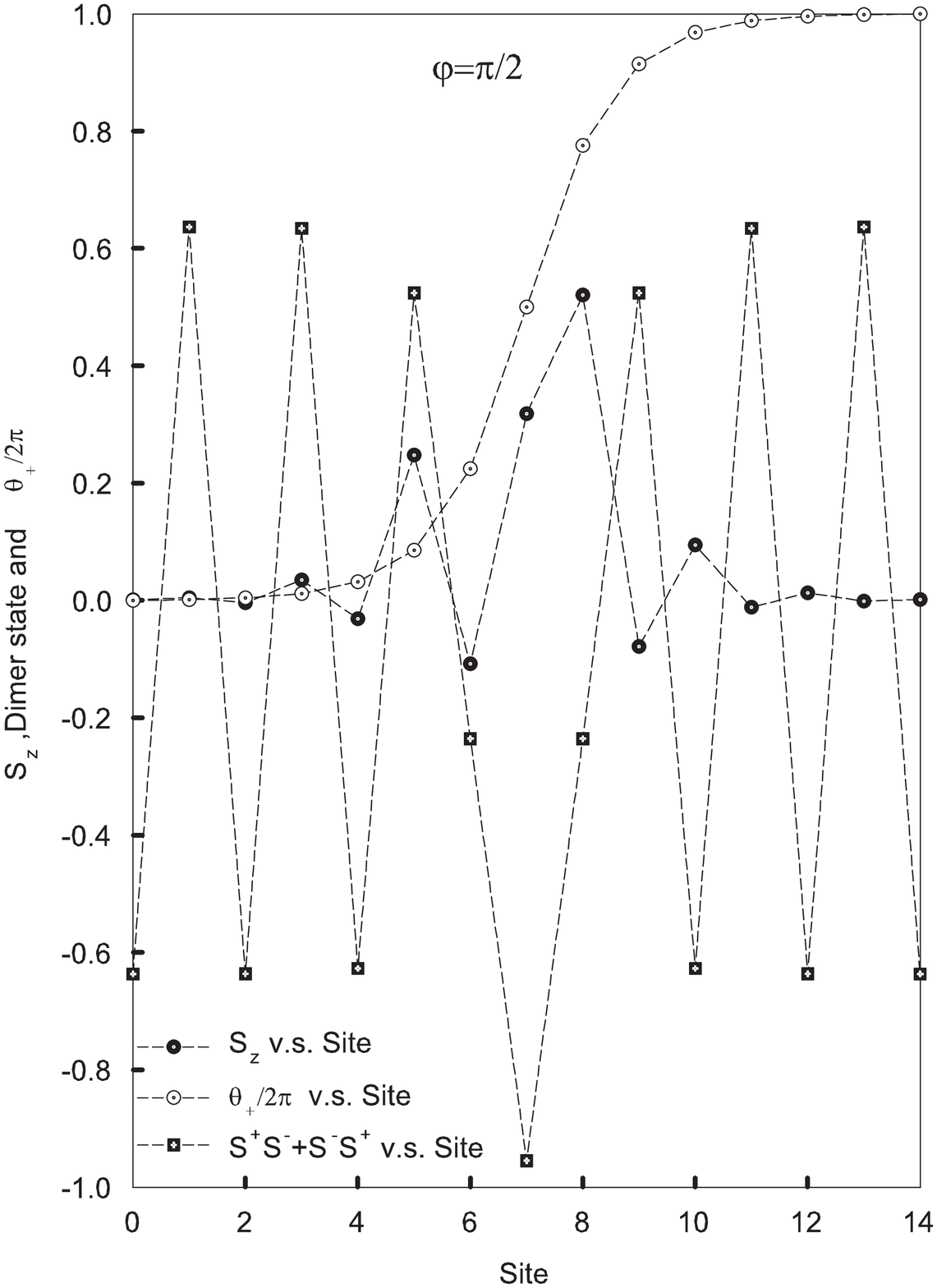}%
\end{center}
\end{figure}
%EndExpansion
\newpage

Table 1

Differential equations satisfied by Jacobian elliptic functions. See Appendix
A for the definations of Jacobian elliptic functions.%

\begin{tabular}
[c]{|l|l|l|l|}\hline
JEF & its equation & JEF & its equation\\\hline
$y=sn(u)$ & $(\partial_{u}y)^{2}=(1-y^{2})(1-k^{2}y^{2})$ & $y=cn(u)$ &
$(\partial_{u}y)^{2}=(1-y^{2})(1-k^{2}+k^{2}y^{2})$\\\hline
$y=dn(u)$ & $(\partial_{u}y)^{2}=(y^{2}-1)(1-k^{2}-y^{2})$ & $y=ns(u)$ &
$(\partial_{u}y)^{2}=(y^{2}-1)(y^{2}-k^{2})$\\\hline
$y=nc(u)$ & $(\partial_{u}y)^{2}=(y^{2}-1)[(1-k^{2})y^{2}+k^{2}]$ & $y=nd(u) $
& $(\partial_{u}y)^{2}=(1-y^{2})[(1-k^{2})y^{2}-1]$\\\hline
$y=sc(u)$ & $(\partial_{u}y)^{2}=(y^{2}+1)(1+k^{\prime2}y^{2})$ & $y=cd(u)$ &
$(\partial_{u}y)^{2}=(y^{2}-1)(k^{2}y^{2}-1)$\\\hline
$y=sd(u)$ & $(\partial_{u}y)^{2}=(1-k^{\prime2}y^{2})(1+k^{2}y^{2})$ &
$y=cs(u)$ & $(\partial_{u}y)^{2}=(1+y^{2})(k^{\prime2}+y^{2})$\\\hline
$y=dc(u)$ & $(\partial_{u}y)^{2}=(k^{2}-y^{2})(1-y^{2})$ & $y=ds(u)$ &
$(\partial_{u}y)^{2}=(y^{2}-k^{\prime2})(y^{2}+k^{2})$\\\hline
\end{tabular}

\end{document}